\documentclass{INTERSPEECH2023}
\usepackage{amsfonts,amssymb}
\usepackage{multirow}
\usepackage{threeparttable}

\interspeechcameraready

\title{CASA-ASR: Context-Aware Speaker-Attributed ASR}
\name{Mohan Shi$^1$, Zhihao Du$^2$, Qian Chen$^2$, Fan Yu$^2$, Yangze Li$^2$, Shiliang Zhang$^2$, Jie Zhang$^1$, Li-Rong Dai$^1$}
\address{
  $^1$NERC-SLIP, University of Science and Technology of China (USTC), China\\
  $^2$Speech Lab of DAMO Academy, Alibaba Group, China}
\email{smohan@mail.ustc.edu.cn}

\begin{document}

\maketitle
 
\begin{abstract}
\vspace{-0.1cm}
Recently, speaker-attributed automatic speech recognition (SA-ASR) has attracted a wide attention, which aims at answering the question ``who spoke what''. Different from modular systems, end-to-end (E2E) SA-ASR minimizes the speaker-dependent recognition errors directly and shows a promising applicability.
In this paper, we propose a context-aware SA-ASR (CASA-ASR) model by enhancing the contextual modeling ability of E2E SA-ASR. Specifically, in CASA-ASR, a contextual text encoder is involved to aggregate the semantic information of the whole utterance, and a context-dependent scorer is employed to model the speaker discriminability by contrasting with speakers in the context.
In addition, a two-pass decoding strategy is further proposed to fully leverage the contextual modeling ability resulting in a better recognition performance. Experimental results on AliMeeting corpus show that the proposed CASA-ASR model outperforms the original E2E SA-ASR system with a relative improvement of 11.76\% in terms of speaker-dependent character error rate.
\end{abstract}

\section{Introduction}
Multi-talker modeling in meeting scenarios, also known as the cocktail party problem, is one of the most challenging tasks in the field of speech signal processing~\cite{fiscus2006rich,fiscus2007rich}. The purpose of speaker-attributed automatic speech recognition (SA-ASR) is not only to obtain a multi-talker transcription~\cite{yu2017recognizing,chang2019mimo,ZhangCQW20,chen2017progressive,kanda2020serialized}, but also to predict the speaker for each character or word in the utterance, that is, to solve the problem of ``who spoke what''. 

To obtain the multi-talker transcriptions, serialized output training (SOT)~\cite{kanda2020serialized} can be used to implement multi-talker ASR, which simply concatenates the transcriptions of different speakers through a special symbol $\langle\text{sc}\rangle$, usually in the order of the starting time of the utterance, known as first-in first-out. Compared with permutation invariant training (PIT)~\cite{yu2017recognizing,chang2019mimo,ZhangCQW20}, where multi-talker transcriptions are obtained through multiple decoders in the model, SOT avoids the limit of the maximum number of speakers and duplicated hypotheses, resulting in a better performance in multi-talker ASR tasks.

In literature, modular~\cite{YuSCSAASR} and end-to-end (E2E)~\cite{KandaGWMCZY20,KandaCGWMCY21SLT,ChangKGWMY21,KandaYGWMCY21} approaches for SA-ASR have been proposed. There are three kinds of modular approaches, including frame-level diarization with SOT (FD-SOT), word-level diarization with SOT (WD-SOT) and target-speaker separation and ASR (TS-ASR). In FD-SOT, the results of frame-level speaker diarization and the hypotheses of SOT-based ASR are aligned to obtain speaker-attributed transcriptions. In WD-SOT, speaker-attributed transcriptions are obtained by word-level speaker diarization on the hypotheses of SOT-based ASR. TS-ASR was proposed to train the joint model of target speaker separation and ASR, which achieves the best SA-ASR performance among the above three kinds of modular approaches. 
In contrast to modular approaches that do not directly optimize the speaker-dependent objective, E2E SA-ASR approaches~\cite{KandaGWMCZY20,KandaCGWMCY21SLT,ChangKGWMY21,KandaYGWMCY21} mainly consists of four modules:  ASR encoder (ASR-Enc), ASR decoder (ASR-Dec), speaker encoder (Spk-Enc) and speaker decoder (Spk-Dec). The ASR-Enc and Spk-Enc obtain the speech representations for multi-talker ASR and speaker identification, respectively. The Spk-Dec mainly takes the token sequence and the representation obtained by Spk-Enc as input and outputs the speaker representation for each token. The posterior probability of each speaker for each token is obtained by cosine similarity scoring between the output of Spk-Dec and speaker profiles. The speaker profile weighted by the posterior probability is fed to the ASR-Dec to help resolve multi-talker ASR transcriptions.

Although the E2E SA-ASR paradigm shows a promising applicability in multi-talker scenarios, the contextual information (that might be beneficial for understanding the acoustic scene of ``who spoke what") was not sufficiently taken into account in existing E2E approaches. This can be interpreted from two aspects. For instance, in~\cite{KandaYGWMCY21} the Spk-Dec takes the token sequence as the query for source-target attention and the speech representation as the key and value to generate the speaker representation for each token. The problem lies in that the raw token sequence lacks context information, which may cause an inaccurate attention range of each speaker and a non-informative speaker representation. Besides, the involved cosine similarity is a kind of context-independent scorer (CI-Scorer) when calculating the posterior probability of each speaker for each token, which is another source of context insufficiency and also lacks nonlinear modeling ability. 

In this work, we therefore propose a context-aware speaker-attributed ASR (CASA-ASR) to enhance the context modeling ability accordingly. We first introduce a contextual text encoder (Context-Enc) to aggregate the semantic information of the whole utterance to obtain a higher quality speaker representation. Then, a context-dependent scorer (CD-Scorer) is employed to model the local speaker discriminability by contrasting with speakers in the context. In addition, we introduce a two-pass decoding to enable Context-Enc and CD-Scorer to obtain more complete context information for better speaker identification performance. Finally, a skip connection is introduced between the raw speaker representation and the output of Spk-Dec to make the speaker information easier to flow into deep representations.
Experimental results in the monaural setting of the Alimeeting corpus~\cite{Yu2022M2MeT,Yu2022Summary}  
 show the superiority of the proposed CASA-ASR method.

\begin{figure}[t!]
 	\centering
 	\includegraphics[width=0.8\linewidth]{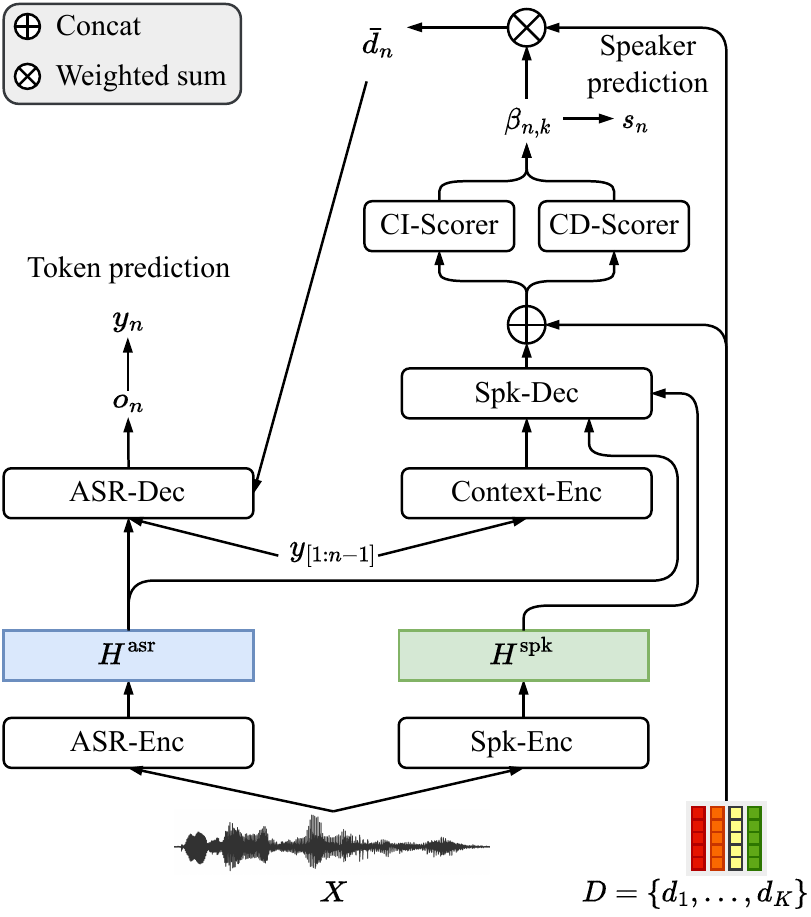}
 	 \vspace{-0.1cm}
 	\caption{
 	 The overall structure of CASA-ASR.
 	}
 	\label{sys_systure_casa-asr}
 	\vspace{-0.4cm}
\end{figure}

\section{Context-aware Speaker-attributed ASR}
\subsection{SOT-based E2E SA-ASR}
\vspace{-0.1cm}
In the framework of SOT, the transcriptions of multiple speakers are connected by a special ``speaker change'' symbol $\langle\text{sc}\rangle$ according to the emission time. Let the acoustic features be denoted by $X$ $\in$ $\mathbb{R}$$^{f^{a}\times{l^{a}}}$ and the set of speaker profiles by $D=\{d_k \in \mathbb{R}^{f^{d}}|k=1,...,K\}$, where $f^{a}$ and $l^{a}$ are the dimension and length of $X$, respectively, $K$ is the total number of speakers in the profile, $d_k$ is the profile of the $k$-th speaker, and $f^{d}$ is the dimension of a profile. The goal of E2E SA-ASR is to estimate the SOT-style transcription $Y=(y_n \in \left\{1,...|\mathcal{V}|\right\}|n=1,...,N)$ and the corresponding speaker of each token $S=(s_n \in \left\{1,...,K\right\}|n=1,...,N)$ with the inputs of $X$ and $D$, where $|\mathcal{V}|$ is the size of vocabulary $\mathcal{V}$ containing the $\langle\text{sc}\rangle$ symbol, $y_n$ and $s_n$ are the word index and speaker index for the $n$-th token, respectively.

\vspace{-0.1cm}
\subsection{CASA-ASR}
\vspace{-0.1cm}

To obtain speaker-attributed transcriptions, E2E SA-ASR mainly includes \textit{ASR branch} and \textit{speaker branch}. The proposed CASA-ASR is obtained by enhancing the context-aware modeling ability of E2E SA-ASR. The overall structure is shown in Figure~\ref{sys_systure_casa-asr}.
\vspace{-0.1cm}
\subsubsection{ASR encoder and speaker encoder}
\vspace{-0.1cm}
We use Conformer~\cite{gulati2020conformer} as ASR encoder (ASR-Enc). Given the acoustic input $X$, the output of ASR-Enc is denoted as ${H}^{\text{asr}}$ $\in$ $\mathbb{R}$$^{f^{h}\times{l^{h}}}$, where $f^{h}$ and $l^{h}$ are the embedding dimension and the length of the sequence, respectively. We use Res2Net~\cite{ZhouZW21} as speaker encoder (Spk-Enc) to transform acoustic features $X$ into speaker embedding $H^{\text{spk}}$ $\in$ $\mathbb{R}$$^{f^{h}\times{l^{h}}}$.

\vspace{-0.2cm}
\subsubsection{ASR decoder}
\vspace{-0.1cm}
The speaker-aware transformer decoder is employed to predict the multi-talker transcriptions, where the weighted speaker profile $\bar{d}_{n}$ $\in$ $\mathbb{R}$$^{f^{d}}$ obtained from the weighted summation of $D$ is added into its first layer and $f^{d}$ represents the dimension of the speaker profile. The calculation procedure of ASR decoder (ASR-Dec) can be summarized as follows:
\vspace{-0.1cm}
\begin{align}
&z_{[1:n-1],1}^{\text{asr}}=\text{PosEnc}(\text{Embed}(y_{[1:n-1]})), \\
&\bar{z}_{n-1,l}^{\text{asr}}=z_{n-1,l}^{\text{asr}}\notag \\
&\quad +\text{MHA}_l^\text{asr-self}(z_{n-1,l}^{\text{asr}},z_{[1:n-1],l}^{\text{asr}},z_{[1:n-1],l}^{\text{asr}}), \\
&\bar{\bar{z}}_{n-1,l}^{\text{asr}}=\bar{z}_{n-1,l}^{\text{asr}}+\text{MHA}_l^\text{asr-src}(\bar{z}_{n-1,l}^{\text{asr}},H^{\text{asr}},H^{\text{asr}}), \\
&z_{n-1,l+1}^{\text{asr}}=
    \begin{cases}
    \bar{\bar{z}}_{n-1,l}^{\text{asr}}+\text{FF}_l^\text{asr}(\bar{\bar{z}}_{n-1,l}^{\text{asr}}+{W^\text{spk}}\cdot{\bar{d}_n})&(l=1) \\
    \bar{\bar{z}}_{n-1,l}^{\text{asr}}+\text{FF}_l^\text{asr}(\bar{\bar{z}}_{n-1,l}^{\text{asr}})&(l>1)
    \end{cases}\\
&o_n=\text{Softmax}({W^o}\cdot{z_{n-1,L^\text{asr}}^\text{asr}}+b^o).
\end{align}
\vspace{-0.1cm}
The token sequence is first processed to obtain $z_{[1:n-1],1}^{\text{asr}}$ $\in$ $\mathbb{R}$$^{f^{d}\times{n-1}}$ by embedding and positional encoding. For each layer $l$ in ASR-Dec, the MHA operation and the source-target- (src-) attention operation are applied at $z_{n-1,l}^{\text{asr}}$ to obtain $\bar{\bar{z}}_{n-1,l}^{\text{asr}}$. Then, the position-wise feed forward layer is applied to obtain the input of the next layer $z_{n-1,l+1}^{\text{asr}}$. When $l=1$, $\bar{d}_{n}$ is projected through a linear transformation by multiplying with $W^\text{spk}$ $\in$ $\mathbb{R}$$^{f^{h}\times{f^{d}}}$. Then, it is added with $\bar{\bar{z}}_{n-1,l}^{\text{asr}}$ and sent to the feed-forward layer. Finally, the outputs of the last decoder layer is passed through a linear transformation and softmax activation function to obtain $o_n$. The posterior probability of token $i$ (i.e., the $i$-th token in $\mathcal{V}$) at the $n$-th decoder step is represented as
\vspace{-0.1cm}
\begin{align}
&P_r(y_n=i | y_{[1:n-1]},s_{[1:n]},X,D)=o_{n,i},
\end{align}
where $o_{n,i}$ represents the $i$-th element of $o_n$.

\vspace{-0.1cm}
\subsubsection{Context-aware speaker decoder}
\vspace{-0.1cm}
Considering that the semantic information of $\bar{z}_{[1:n-1],1}^{\text{asr}}$ is relatively weak, in order to obtain a better context-aware speaker representation, we apply a contextual text encoder (Context-Enc) to $\bar{z}_{[1:n-1],1}^{\text{asr}}$ to obtain $\bar{z}_{[1:n-1]}^{\text{context}}$. The first layer of Spk-Dec with Context-Enc can be represented as follows:
\vspace{-0.1cm}
\begin{align}
&\bar{z}_{n-1}^{\text{context}}=\text{Context-Enc}(\bar{z}_{[1:n-1],1}^{\text{asr}}), \\
&\bar{\bar{z}}_{n-1,1}^{\text{spk}}=\text{MHA}_{n-1,1}^{\text{spk-src}}(\bar{z}_{n-1}^{\text{context}},H^{\text{asr}},H^{\text{spk}}), \\
&z_{n-1,2}^{\text{spk}}=\bar{\bar{z}}_{n-1,1}^{\text{spk}}+\text{FF}_1^\text{spk}(\bar{\bar{z}}_{n-1,1}^{\text{spk}}).
\end{align}
The Context-Enc consists of several layers of transformer encoder containing self-attention and feed forward layers. By treating $\bar{z}_{n-1}^{\text{context}}$ as the query, $H^{\text{asr}}$ as the key and $H^{\text{spk}}$ as the value, a context-dependent speaker representation $\bar{\bar{z}}_{n-1,1}^{\text{spk}}$ are obtained for each token through the src-attention. 
Since $\bar{z}_{[1:n-1]}^{\text{context}}$ contains rich contextual information, it can aggregate more speaker embeddings $H^{\text{spk}}$ by interacting with $H^{\text{asr}}$, 
which leads to a higher-quality speaker representation for each token. We employ a standard transformer-based decoder consisting of self-attention, source-attention and point-wise feed forward layers as the remaining layer of Spk-Dec to predict the speaker representation for each token, which is computed as:
\vspace{-0.1cm}
\begin{align}
&\bar{z}_{n-1,l}^{\text{spk}}=z_{n-1,l}^{\text{spk}}\notag \\
&\quad +\text{MHA}_l^\text{spk-self}(z_{n-1,l}^{\text{spk}},z_{[1:n-1],l}^{\text{spk}},z_{[1:n-1],l}^{\text{spk}}), \\
&\bar{\bar{z}}_{n-1,l}^{\text{spk}}=\bar{z}_{n-1,l}^{\text{spk}}+\text{MHA}_l^\text{spk-src}(\bar{z}_{n-1,l}^{\text{spk}},H^{\text{spk}},H^{\text{spk}}), \\
&z_{n-1,l+1}^{\text{spk}}=\bar{\bar{z}}_{n-1,l}^{\text{spk}}+\text{FF}_l^\text{spk}(\bar{\bar{z}}_{n-1,l}^{\text{spk}}), \\
&q_n={W^q}\cdot({z_{n-1,L^\text{spk}}^\text{spk}}+\bar{\bar{z}}_{n-1,1}^{\text{spk}}).
\end{align}
Since the Spk-Enc is pre-trained, the initial speaker representation from the first layer of Spk-Dec is also meaningful. Therefore, we add a skip connection from the first layer to the output layer to make the speaker information easier to flow into deep representations, i.e., the speaker representation for each token $\bar{\bar{z}}_{n-1,1}^{\text{spk}}$ is added to the $L^\text{spk}$-th layer output. Finally, the output $q_n$ is obtained by linear transformation.
\vspace{-0.1cm}
\subsubsection{CI-CD Scorer}
\vspace{-0.1cm}
In E2E SA-ASR~\cite{KandaYGWMCY21}, the output of Spk-Dec $q_n$ and the $k$-th speaker vector $d_k$ are computed by cosine similarity to obtain the weight $\beta_{n,k}$ of $k$-th speaker for $n$-th token. However, the cosine similarity is a context-independent scorer (CI-Scorer), which lacks context information and nonlinear modeling ability. Inspired by \cite{SOND}, we employ an extra context-dependent scorer (CD-Scorer) to enhance the context-aware and nonlinear modeling ability, resulting in the proposed CI-CD Scorer:
\vspace{-0.1cm}
\begin{align}
&score_{n,k}^\text{CI}=\text{cos}(q_n,d_k), \\
&score_{n,k}^\text{CD}=\text{tanh}(\text{CD-Scorer}(q_{[1:n]},d_k)), \\
&\beta_{n,k}=\frac{\text{exp}(score_{n,k}^\text{CI}+score_{n,k}^\text{CD})}{\textstyle\sum_j^K\text{exp}(score_{n,j}^\text{CI}+score_{n,j}^\text{CD})}, \label{eq16} \\
&\bar{d}_n=\sum_{k=1}^K\beta_{n,k}d_k.
\end{align}
where $\text{cos}$ means the cosine similarity. The concatenation of $q_{[1:n]}$ $\in$ $\mathbb{R}$$^{f^{d}\times{n}}$ with $d_k$ $\in$ $\mathbb{R}$$^{f^{d}}$ is used as the input of $\text{CD-Scorer}$, which consists of several transformer encoder layers. The output of $\text{CD-Scorer}$ is passed through a $\text{tanh}$ function to obtain $score_{n,k}^\text{CD}$, which matches the range of [-1, 1] as $score_{n,k}^\text{CI}$. Then, $score_{n,k}^\text{CI}$ and $score_{n,k}^\text{CD}$ are added and normalized over $K$ speakers to get $\beta_{n,k}$ $\in$ $\mathbb{R}$ as in (\ref{eq16}). Finally, $\bar{d}_n$ is obtained by the weighted sum of $d_k$ and $\beta_{n,k}$. The posterior probability of person $k$ speaking the $n$-th token is given by
\vspace{-0.1cm}
\begin{align}
\beta_{n,k}=P_r(s_n=k | y_{[1:n-1]},s_{[1:n-1]},X,D).
\end{align}
The entire model is trained as a joint task using the loss function:
\begin{align}
&Loss^\text{joint}=\lambda{Loss^\text{spk}(\hat{s},s)}+(1-\lambda){Loss^\text{asr}(\hat{y},y)}
\end{align}
where $s$ and $y$ are the true speaker and transcription, and $\hat{s}$ and $\hat{y}$ are the corresponding hypothesises, respectively. The cross-entropy function is used as the speaker loss while joint attention-CTC loss~\cite{GravesFGS06,WatanabeHKHH17} is used to train the ASR model.
\vspace{-0.1cm}
\subsubsection{Two-Pass Decoding}
\vspace{-0.1cm}
Context-Enc and CD-Scorer need the whole sequence to achieve a better context modeling. However, at each step of decoding, the input of the Context-Enc and CD-Scorer is the partial sequence obtained from the previous steps, i.e., $\bar{z}_{[1:n-1],1}^{\text{asr}}$, $q_{[1:n]}$), which leads to insufficient global context information. Therefore, after the best hypothesis of ASR $y_{[1:N]}$ is obtained in the first-pass decoding, we take $y_{[1:N]}$ as the input for the second-pass decoding, which is only used to predict speakers. In the second-pass decoding, the input of Context-Enc and CD-Scorer has global context, i.e., $\bar{z}_{[1:N],1}^{\text{asr}}$, $q_{[1:N]}$), which can make the speaker identification more accurate. Note that as in the second-pass decoding the input of Context-Enc and CD-Scorer is complete, the speaker prediction can therefore be obtained in one feed-forward without any auto-regression. With the two-pass decoding, the posterior probability of person $k$ speaking the $n$-th token is thus given by:
\begin{align}
\beta_{n,k}^\text{two-pass}=P_r(s_n^\text{two-pass}=k | y_{[1:N]},s_{[1:N]},X,D).
\end{align}

\section{Experiments}
\subsection{Dataset and evaluation metrics}
\vspace{-0.1cm}
In this work, Alimeeting corpus is used to evaluate the performance, which is an open source Mandarin corpus of meeting scenario. This dataset contains 104.75 hours for training (Train), 4 hours for evaluation (Eval) and 10 hours for test (Test). 
Among them, the Train and Eval sets contain not only the 8-channel far-field audios recorded by a microphone array (\textit{Ali-far}), but also the single-channel near-field audios recorded by the headset microphone of the participants (\textit{Ali-near}), while the Test set only contains the far-field audios. Because a monaural setup is used for training and evaluation in this work, we apply CDDMA beamforming~\cite{huang2020differential,ZhengHWSFY21} to \textit{Ali-far} to obtain a single-channel data set \textit{Ali-far-bf}. The prefixes \textit{Train-}, \textit{Eval-} and \textit{Test-} stand for different sets. For the speaker profile, we use a 256-dim d-vector (i.e., $f^{d}$ = 256) which is extracted by Res2Net pre-trained on CN-Celeb corpus~\cite{FanKLLCCZZCW20}. Since there are at most four participants in each scenario, in case of meetings with less participants, the profiles are padded with speakers from other meetings.

We use speaker independent- (SI-) and speaker dependent- (SD-) character error rate (CER)~\cite{fu2021aishell} as evaluation metrics. SI-CER is used to evaluate the multi-talker ASR task in the SOT framework, which is computed in the same way as the normal CER, ignoring the speaker labels. While SD-CER is calculated by matching the ASR hypothesis to the corresponding speaker reference transcription, which is a rigorous evaluation metric used to evaluate SA-ASR in meeting scenarios.

\begin{table*}[t]
\renewcommand{\thetable}{2}
\centering
\caption{Ablation study on different modules of the proposed CASA-ASR in terms of SI-CER (\%) and SD-CER (\%).}
\vspace{-0.25cm}
\setlength{\tabcolsep}{3mm}
\begin{tabular}{l|c|ccc|ccc}
\toprule
\hline
\multicolumn{1}{c|}{\multirow{2}{*}{Approach}} & \multicolumn{1}{c|}{\multirow{2}{*}{Parameters}} & \multicolumn{3}{c|}{SI-CER} & \multicolumn{3}{c}{SD-CER} \\ \cline{3-8}
\multicolumn{1}{c|}{} & \multicolumn{1}{c|}{} & Eval & Test & Average & Eval & Test & Average \\ \hline
E2E SA-ASR & 60.07 M & 26.4 & 28.1 & 27.6 & 31.8 & 34.7 & 34.0 \\  \hline
\quad + skip connection & 60.07 M & 26.5 & 28.1 & 27.7 & 30.4 & 34.7 & 33.6 \\ 
\quad \quad + CD-Scorer & 65.46 M & 26.5 & 28.1 & 27.7 & 29.7 & 32.5 & 31.8 \\ 
\quad \quad \quad + two-pass decoding & 65.46 M & 26.5 & 28.1 & 27.7 & 28.7 & 31.4 & 30.7 \\
\quad \quad \quad \quad + Context-Enc (CASA-ASR) & 70.79 M & \textbf{26.3} & \textbf{28.0} & \textbf{27.6} & \textbf{27.9} & \textbf{30.8} & \textbf{30.0} \\ \hline
\bottomrule
\end{tabular}
\label{tab:ablation_study}
\vspace{-0.1cm}
\end{table*}

\begin{table*}[t]
\renewcommand{\thetable}{3}
\centering
\caption{Comparison of recognition performance on different data settings in terms of SI-CER (\%) and SD-CER (\%).}
\vspace{-0.25cm}
\setlength{\tabcolsep}{3mm}
\begin{tabular}{l|c|ccc|ccc}
\toprule
\hline
\multicolumn{1}{c|}{\multirow{2}{*}{Data Selection}} & \multicolumn{1}{c|}{\multirow{2}{*}{Duration}} & \multicolumn{3}{c|}{SI-CER} & \multicolumn{3}{c}{SD-CER} \\ \cline{3-8}
\multicolumn{1}{c|}{} & \multicolumn{1}{c|}{} & Eval & Test & Average & Eval & Test & Average \\ \hline
\textit{Train-Ali-far-bf} & 105 Hours & 27.1 & 28.6 & 28.2 & 29.9 & 32.2 & 31.6 \\ 
\quad + \textit{Train-Ali-near} & 210 Hours & 26.4 & 28.2 & 27.7 & 29.0 & 32.2 & 31.4 \\
\quad \quad + interference for \textit{Train-Ali-near} & 210 Hours & 26.4 & 28.2 & 27.8 & 28.5 & 30.8 & 30.2 \\
\quad \quad \quad + interference for \textit{Train-Ali-far-bf} & 210 Hours & \textbf{26.3} & \textbf{28.0} & \textbf{27.6} & \textbf{27.9} & \textbf{30.8} & \textbf{30.0} \\ \hline
\bottomrule
\end{tabular}
\label{tab:data_selection}
\vspace{-0.3cm}
\end{table*}

\vspace{-0.1cm}
\subsection{Model configuration}
\vspace{-0.1cm}
The ASR-Enc contains 12 layers of conformer with 4-head multi-head attention (MHA) and convolutional kernel size of 15, where the dimensions of MHA and feed-forward network (FFN) are set to be 256 (i.e., $f^{h}$ = 256) and 2048, respectively. The ASR-Dec contains 6 layers of transformer decoder. The Spk-Enc has the same architecture as the d-vector extractor except that an additional linear layer is added to map the $f^{d}$-dim output into $f^{h}$-dim embedding. Spk-Enc is initialized with the pretrained d-vector extractor. The Spk-Dec has 3 layers, i.e., $L^\text{spk}$ = 3. Both Context-Enc and CD-Scorer consist of 4 transformer encoder layers with 4-head MHA. The settings of MHA and FFN modules are similar to ASR-Enc. The weight of the additional CTC loss in \cite{WatanabeHKHH17} is set to be 0.3. 
\vspace{-0.1cm}
\subsection{Training details}
\vspace{-0.1cm}
In this work, we use the 80-dimensional log Mel filterbank (Fbank) as the input feature. The window size is 32 ms and the window shift is 8 ms. The vocabulary used for ASR consists of 4950 common Chinese characters ($|\mathcal{V}|$ = 4950). We use the ESPnet toolkit~\cite{watanabe2018espnet} to train the models.

First, we train a SOT-based multi-talker ASR model under the above configurations using \textit{Train-Ali-far-bf} and \textit{Train-Ali-near}. Then, we use the pre-trained parameters of the ASR model and d-vector extractor to initialize ASR-Enc, ASR-Dec, and Spk-Enc. 
Finally, we jointly fine-tune the model using \textit{Train-Ali-far-bf} and \textit{Train-Ali-near}, where the weight $\lambda$ for speaker loss is set to 0.5. The total amount of data used in the experiment is consistent with that in~\cite{YuSCSAASR} to ensure fair comparison.
We used the Adam~\cite{2014Adam} optimizer to jointly train the model for 60 epochs with a warmup for 2000 steps, and the maximum learning rate is set to 5e-4. 
The speaker with the highest averaged $\beta_{n,k}$ of tokens between two $\langle\text{sc}\rangle$ is selected as the predicted speaker of that utterance.

\begin{table}[h!]
\renewcommand{\thetable}{1}
\centering
\caption{Comparison of various approaches on AliMeeting sets in terms of SD-CER (\%).}
\vspace{-0.25cm}
\setlength{\tabcolsep}{4mm}
\begin{tabular}{c|ccc}
\toprule
\hline
\multirow{2}{*}{Approach} & \multicolumn{3}{c}{SD-CER} \\ \cline{2-4}
\multicolumn{1}{c|}{} & Eval & Test & Average \\ \hline
FD-SOT~\cite{YuSCSAASR} & 41.0 & 41.2 & 41.2 \\ 
WD-SOT~\cite{YuSCSAASR} & 36.0 & 37.1 & 36.8 \\ 
TS-ASR~\cite{YuSCSAASR} & 32.5 & 35.1 & 34.4 \\ \hline
E2E SA-ASR \tnote{$^*$}~\cite{KandaYGWMCY21} & 31.8 & 34.7 & 34.0 \\ 
CASA-ASR (ours) & \textbf{27.9} & \textbf{30.8} & \textbf{30.0} \\ \hline
\bottomrule
\end{tabular}
\begin{tablenotes}
    \footnotesize
		\item $^*$: This models is re-implemented by ourselves.
\end{tablenotes}
\label{tab:result_pipline}
\vspace{-0.3cm}
\end{table}

\subsection{Experimental Results}
\subsubsection{Comparison of various SA-ASR approaches}
\vspace{-0.1cm}
Table~\ref{tab:result_pipline} shows the comparison of the proposed CASA-ASR, three modular approaches (i.e., FD-SOT, WD-SOT, TS-ASR) and the E2E SA-ASR. As expected, the E2E SA-ASR slightly outperforms the best modular approach, TS-ASR, in terms of the average SD-CER on Eval and Test sets (from 34.4\% to 34.0\%). The proposed method CASA-ASR achieves 11.8\% (from 34.0\% to 30.0\%) relative SD-CER reduction compared to E2E SA-ASR.
\vspace{-0.1cm}
\subsubsection{The ablation study of CASA-ASR}
\begin{figure}[tb]
 	\centering
 	\includegraphics[width=0.8\linewidth]{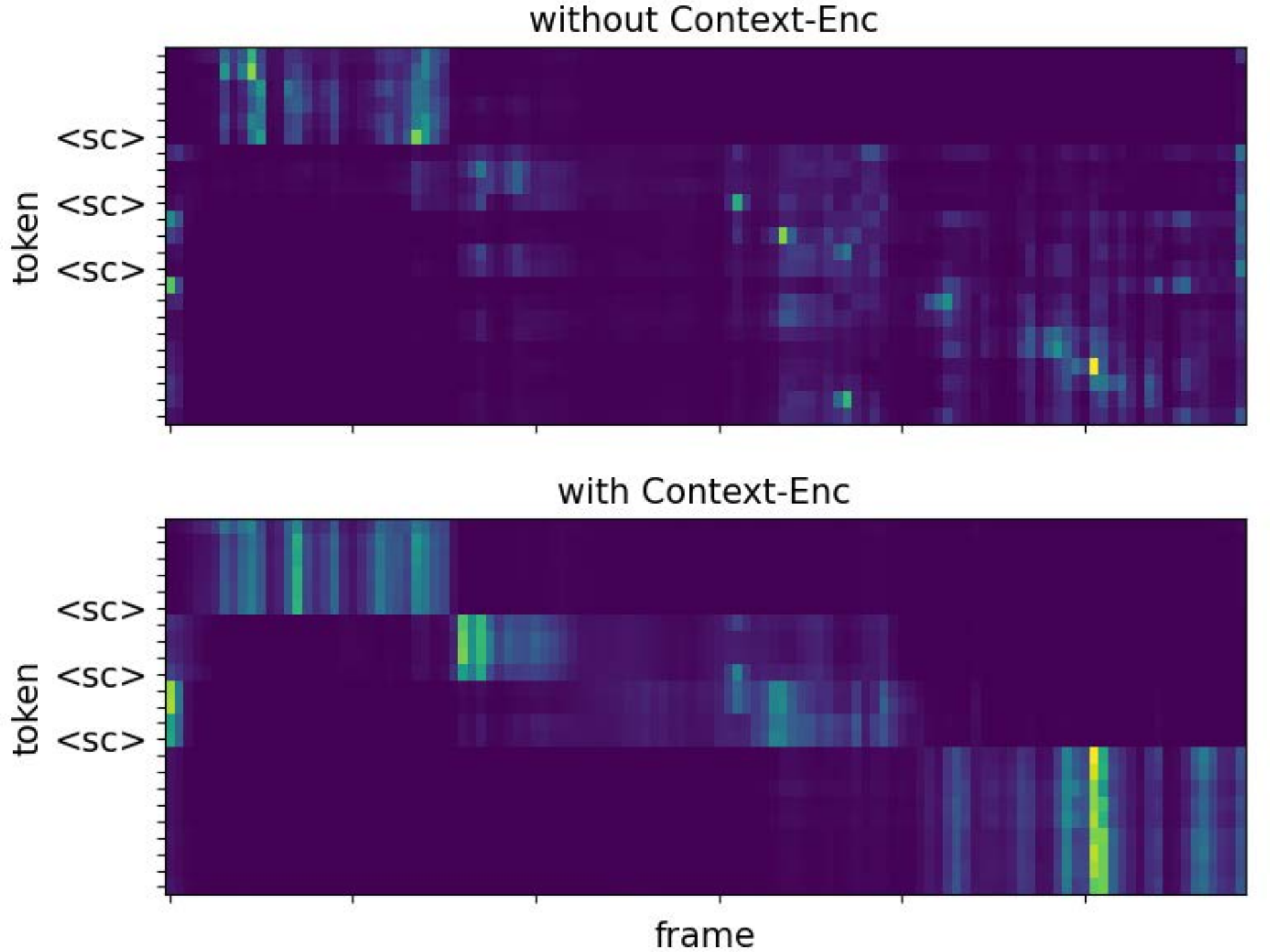}
 	 \vspace{-0.2cm}
 	\caption{
 	 Visualization of source-target attention with and without Context-Enc.
 	}
 	\label{visualize}
 	\vspace{-0.7cm}
\end{figure}

\vspace{-0.1cm}
We show the ablation study on different modules of the proposed CASA-ASR in Tabel~\ref{tab:ablation_study}. Skip connection is applied to assist speaker information flow into deep representations, leading to a slight SD-CER reduction (from 34.0\% to 33.6\%) on Eval and Test sets. Involving the CD-Scorer leads to a large relative SD-CER reduction (from 33.6\% to 31.8\%), which indicates that CD-Scorer can improve the performance of SA-ASR by contrasting with other speakers in the context. Due to the properties of autogressive decoding, the context of each step is incomplete when decoding. The two-pass decoding is involved to solve this problem, which further reduces the average SD-CER (from 31.8\% to 30.7\%). The Context-Enc is involved to obtain the final CASA-ASR, which achieves the best performance in terms of both SD-CER (30.0\%) and SI-CER (27.6\%).
As shown in Fig.~\ref{visualize}, by involving the Context-Enc, the tokens (query for src-attention) between two speaker change symbols $\langle\text{sc}\rangle$ have a more focused attention weights for the speech representation. This indicates that the Context-Enc can aggregate the context information of the whole sentence to obtain a higher-quality speaker representation through source-target attention.
\vspace{-0.4cm}
\subsubsection{Comparison of performance on different data settings}
\vspace{-0.1cm}
We also investigate the SA-ASR performance under different data selection and the results are shown in Tabel~\ref{tab:data_selection}. In case of only training with \textit{Train-Ali-far-bf}, we achieve an average SI-CER of 28.2\% and SD-CER of 31.6\% on Eval and Test sets. Simply adding \textit{Train-Ali-near} to the training set can improve the performance in terms of SI-CER (from 28.2\% to 27.7\%). However, it can not take a substantial improvement for SD-CER because there is only one speaker in each audio in \textit{Train-Ali-near}. After adding interference speakers to \textit{Train-Ali-near}, the average SD-CER is significantly reduced (from 31.4\% to 30.2\%). Additionally adding the interference speaker to the \textit{Train-Ali-far-bf} achieves the lowest SD-CER (30.0\%).

\vspace{-0.15cm}
\section{Conclusion}
\vspace{-0.1cm}
In this paper, we proposed a context-aware speaker-attributed ASR model by enhancing the contextual modeling capability of E2E speaker-attributed ASR. It was shown that adding skip connection can make the speaker information easier to flow into deep representations. Then Context-Enc enhances the context-aware modeling and obtains a higher-quality speaker representation for each token. With the CD-Scorer, stronger context awareness and nonlinear modeling ability are involved into the model.
The proposed two-pass decoding strategy makes the context of Context-Enc and CD-Scorer more complete, which further improves performance. The proposed CASA-ASR approach achieves the best performance on the open-source Alimeeting corpus.


\end{document}